\documentclass[prb,superscriptaddress,twocolumn,showpacs]{revtex4}

\usepackage{graphicx}
\input{epsf}
\usepackage{multirow}

\begin{document}

\title{Infrared spectroscopic studies on unoriented single-walled carbon nanotube 
films under hydrostatic pressure}

\author{K. Thirunavukkuarasu}
\affiliation{Experimentalphysik II, Universit\"at Augsburg,
Universit\"atsstr.\ 1, 86159 Augsburg, Germany}

\author{F. Hennrich}
\affiliation{Institut f\"ur Nanotechnologie, Forschungszentrum
Karlsruhe, 76021 Karlsruhe, Germany}
\author{K. Kamar\'as}
\affiliation{Research Institute for Solid State
Physics and Optics, 1525 Budapest, Hungary}
\author{C. A. Kuntscher}\email{christine.kuntscher@physik.uni-augsburg.de}
\affiliation{Experimentalphysik II, Universit\"at Augsburg,
Universit\"atsstr.\ 1, 86159 Augsburg, Germany}

\date{\today}

\begin{abstract}
The electronic properties of as-prepared and 
purified unoriented single-walled carbon nanotube films were studied
by transmission measurements over a broad frequency range (far-infrared up 
to visible) as a function of temperature (15 K - 295 K) and external pressure 
(up to 8~GPa). Both the as-prepared and the purified SWCNT films exhibit nearly 
temperature-independent properties. 
With increasing pressure the low-energy absorbance decreases suggesting an
increasing carrier localization due to pressure-induced deformations.
The energy of the optical transitions in the SWCNTs decreases with increasing
pressure, which can be attributed to pressure-induced hybridization and
symmetry-breaking effects. We find an anomaly in the
pressure-induced shift of the optical transitions at $\sim$2~GPa due to a structural
phase transition.
\end{abstract}


\maketitle

\section{\label{sec:intro}Introduction}

Carbon nanotubes have been intensely studied since their discovery in 1991. In particular, 
single-walled carbon nanotubes (SWCNTs) raised great interest, since they can be considered
as the closest realization of a one-dimensional system
due to their unique structure and properties, \cite{Dresselhaus96} which strongly depend
on the nanotubes' geometrical parameters like diameter and chirality.\cite{Reich04}
An understanding of the structural and electronic properties of carbon nanotubes
is of general importance, and one efficient approach is to study the changes
associated with the variation of the environmental parameters. 
Here we expose SWCNTs to extreme conditions, i.e, low temperature and high pressure, 
and study their frequency-dependent optical response.

Both the low-energy electronic properties and the optical band gaps of SWCNTs
are expected to show a significant temperature dependence, determined
by the diameter and chirality of the
nanotubes.\cite{Lefebvre04, Capaz05, Cronin06}
It is thus very surprising that only a very weak temperature dependence of the low-energy properties
was observed by spectroscopic investigations,\cite{Ugawa99,Ugawa01, Hilt00, Hone98, Borondics06}
and it was attributed to the localization of charge
carriers.\cite{Hilt00, Benoit02, Fuhrer99, Jiang01} 
In contrast, several experimental studies showed that the application of external pressure 
has drastic effects on SWCNTs. Despite some contradictory experimental 
results, the generally accepted picture is that the pressurized SWCNTs undergo 
a structural phase transition at a critical pressure, where the tubes change
their shape. The nature of the pressure-induced structural deformation of
carbon nanotubes was probed by Raman spectroscopy,\cite{Loa03, Lebedkin06} x-ray
scattering,\cite{Tang00, Sharma01} and neutron diffraction\cite{Rols01}.
The wide-spread results of the various 
experimental studies may arise from differences in the samples' synthesis
and chemical processing, like purification. Furthermore, the role of the
pressure-transmitting medium in the pressure studies is not yet clear.

Many spectroscopic techniques like Raman and scanning tunneling 
spectroscopy were extensively used to understand the complex physics of carbon
nanotubes. Infrared spectroscopy is a very useful tool that
investigates the electronic properties of materials over a broad
energy range close to the Fermi level. Thus, it provides valuable
information on the small-gap tube absorption and the optical transitions
between the Van Hove singularities at higher energies.
Some investigations on nanotubes using infrared spectroscopy have been performed, 
in order to study the influence of temperature and 
doping on their electronic properties.\cite{Bommeli97, Ugawa99, Itkis02, Itkis03, Borondics06} 
However, absorbance measurements under pressure were carried out 
so far only rarely. An early study
of the optical absorption features in SWCNTs under high pressure
employed a less hydrostatic (i.e., solid) pressure transmitting
medium;\cite{Kazaoui00} the broadening and disappearance of the
optical absorption bands were observed for pressures at around
4~GPa. A more recent pressure-dependent investigation on debundled
SWCNTs in suspensions was limited to very low pressures
($<$2.0~GPa) due to the solidification of the aqueous solution at
higher pressures \cite{Wu04}. Thus, the interesting effects expected
under hydrostatic conditions and at high pressure were not studied
by infrared spectroscopy up to now.

In this paper we present the electronic properties of SWCNTs as a function of temperature
and pressure studied by infrared spectroscopy. The investigations were carried out on 
as-prepared and purified unoriented SWCNT thin films over a broad frequency range. 
We show that the temperature variation and the application of pressure affect 
the electronic properties of the SWCNTs very differently. While the low-energy electronic
properties of the SWCNTs are nearly independent of temperature,
they undergo significant changes with the application of pressure.
Furthermore, we clearly find the signature of a pressure-induced 
structural phase transition. 

The paper is organized as follows: The experimental details are presented in 
Sec.\ \ref{Experiment} and the studied materials are described in 
Sec.\ \ref{Materials}. In Sec.\ \ref{Results} we present the results
obtained at ambient conditions (Sec.\ \ref{sec:comparison}), as a function
of temperature (Sec.\ \ref{sec:T-dependence}), and as a function of pressure 
(Sec.\ \ref{sec:P-dependence}). Sec.\ \ref{Results} also includes the analysis of
the results. We discuss the implications of our findings in Sec.\ \ref{sec:discussion}
and summarize our findings in Sec.\ \ref{sec:conclusions}.

\section{Experiment} \label{Experiment}

Temperature-dependent transmission measurements on 
as-prepared and purified SWCNT films were performed over a
broad frequency range (30~cm$^{-1}$ - 22000~cm$^{-1}$) for the
temperatures 15~K - 295~K, using a Bruker IFS66v/S Fouriertransform
infrared (FTIR) spectrometer in combination with
a continuous-flow cold finger cryostat. The room
temperature data in the THz frequency range (6~cm$^{-1}$ -
30~cm$^{-1}$) were obtained with a coherent source
millimeter-submillimeter wave spectrometer.

\begin{figure}
\includegraphics[width=1\columnwidth]{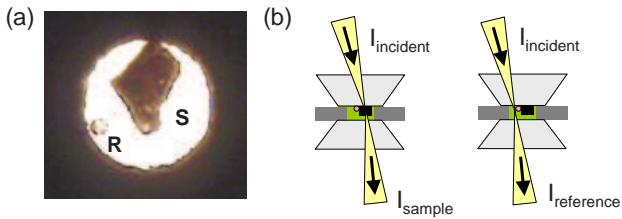}
\caption{\label{fig:loadview} (a) A typical view of the pressure
cell loaded with a piece of carbon nanotube film marked as sample
(S) together with the ruby ball (R) for pressure determination.
The cell is filled with argon as pressure-transmitting medium. (b)
An illustration of a transmission measurement configuration in a
diamond anvil cell.}
\end{figure}

Pressure-dependent transmission measurements at room temperature
were performed in the frequency range from 120~cm$^{-1}$ to
20000~cm$^{-1}$ using a Bruker IFS66v/S FTIR spectrometer combined
with an infrared microscope (Bruker IRscope II). The pressure (up to 8 GPa) was
generated by a diamond anvil cell (DAC) of Syassen-Holzapfel
type\cite{Huber77} for the mid-infrared (MIR) to visible frequency
range. The far-infrared (FIR) measurements were performed using a clamp diamond
anvil cell (Diacell Cryo DAC Mega). The ruby
luminescence method was used for the pressure determination
\cite{mao86}. The typical size of the piece of nanotube film used
in one pressure experiment was approximately 100~$\mu$$m$$\times$100~$\mu$$m$ for the
MIR to visible frequency range and 250~$\mu$$m$$\times$250~$\mu$$m$ for
the FIR frequency range. Liquid argon served as pressure transmitting medium
to ensure hydrostatic conditions. A typical loading of the DAC 
with a piece of carbon nanotube film is shown in Fig.\
\ref{fig:loadview}(a). The piece of the nanotube film is seen in
the transmission mode together with a small ruby ball used for the
pressure determination. Fig. \ref{fig:loadview}(b) illustrates the
configuration of a transmission measurement through a DAC. The
pressure-dependent transmission spectra were recorded by taking
the ratio of intensity transmitted through the sample placed in
the cell and the intensity of radiation transmitted through the argon
(pressure transmitting medium), i.e., the transmittance is given as
T=I$_{sample}$/I$_{reference}$ and the absorbance is calculated according to
A=-log$_{10}$T.

\begin{figure}
\includegraphics[scale=0.4]{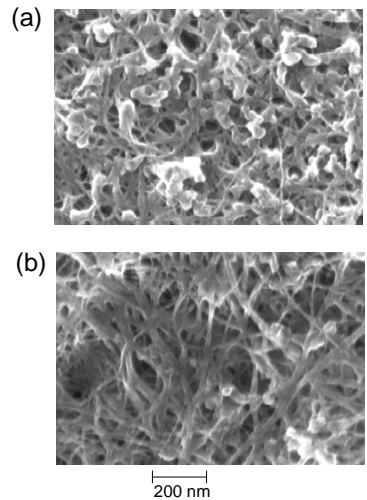}
\caption{\label{fig:SEMimages} Scanning electron microscopy
images of (a) the as-prepared unoriented SWCNT film and (b) the
purified unoriented SWCNT film.}
\end{figure}

\section{Materials}  \label{Materials}

The SWCNTs were prepared by the laser ablation technique using 1:1 Ni/Co
catalyst, and thin films of the SWCNTs were obtained by vacuum
filtration.\cite{Hennrich02,Hennrich03} The average diameter of
the nanotubes is 1.2-1.4~nm. The average density of the nanotubes
in the studied films were found to be
1.2$\pm$0.1~gcm$^{-3}$.\cite{Hennrich02} The purified nanotubes
were obtained by treating them for 48h in HNO$_3$ acid
reflux.\cite{Hennrich02} The purified and as-prepared nanotubes
were then suspended in dimethyl-formamide (DMF) and sonicated
before vacuum filtration.\cite{Hennrich02} The purified thin films
were finallyannealed to remove effects of acid treatment.

Due to the described treatment, we expect defects and shortenings 
for both purified and as-prepared SWCNTs .\cite{Dillon99,Hu03,Monthioux01,Zhang03}
Thus, the adsorption of argon 
used here as pressure transmitting medium is expected to be important and 
needs to be considered. Adsorption of argon occurs at interstitial channels
between the adjacent tubes, grooves and surface sites at the outer
surface of the bundle, and the adsorption sites inside the empty
nanotubes.\cite{Rols05} As the studied nanotube films have not
been subjected to the degassing procedure, it is not possible to
estimate the amount of adsorbed argon, which is even otherwise a
difficult task.

The scanning electron microscopy (SEM) images of the as-prepared
and purified SWCNT films (Fig.\ \ref{fig:SEMimages}) were recorded using 
a digital field emission scanning electron microscope (LEO 982) 
with magnification of 10x to 700000x. They illustrate the
morphological differences between the two studied samples: The
as-prepared film contains more (catalyst) impurities when compared
to the purified film. From the SEM images, a rough estimation of
the bundle diameters in both films could be made. The as-prepared
and the purified films have bundle diameters in the range
6-25~nm and 12-40~nm, respectively. The larger bundle
diameters observed in the SEM images of the purified film is
similar to that observed by Martinez \emph{et
al}.\cite{Martinez03} Both the as-prepared and
the purified SWCNTs have small structural defects, with the purified
film containing smaller amounts of non-carbonaceous impurities.

\begin{figure}
\includegraphics[width=1\columnwidth]{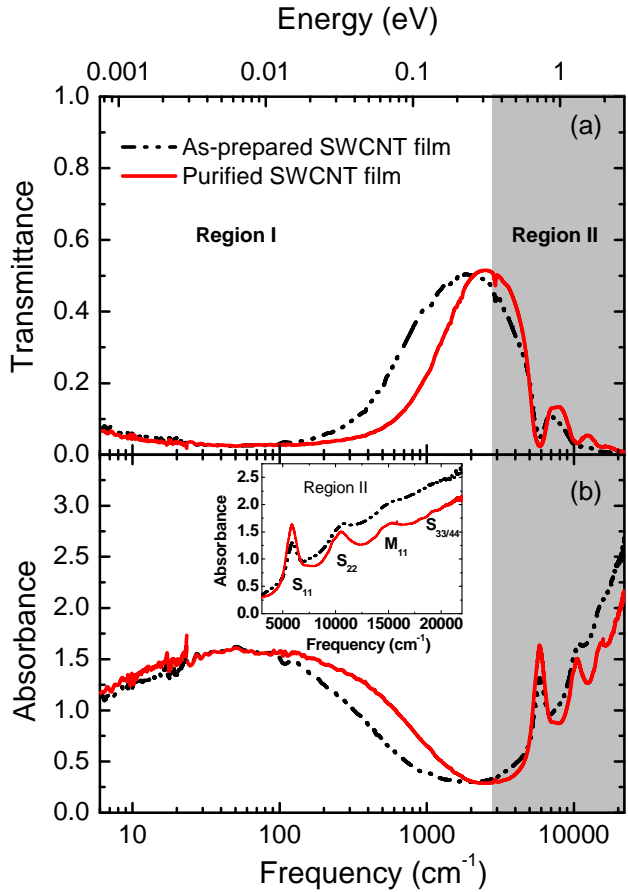}
\caption{\label{fig:Ambientspectra} (a) Transmittance and (b)
absorbance spectra of the as-prepared and the purified unoriented
SWCNT film at ambient conditions. Region I is the frequency range
of low-energy excitations, while Region II (shaded area)
corresponds to the frequency range of the optical
transitions. The inset shows the various contributions of
semiconducting (marked as S) and metallic (marked as M) SWCNTs in
Region II.}
\end{figure}

\section{Results and Analysis} \label{Results}

\subsection{\label{sec:comparison} Ambient conditions}

The transmittance and absorbance spectra of the as-prepared
SWCNT film and purified SWCNT film are shown in Fig.\
\ref{fig:Ambientspectra}. The spectra can be split into two main
energy regions: Region I dominated by a strong FIR absorption band
and Region II (shaded area) with several absorption bands
labelled, S$_{11}$, S$_{22}$, M$_{11}$, and S$_{33/44}$
(see inset of Fig. \ref{fig:Ambientspectra}), which correspond
to interband transitions between the Van Hove
singularities in the density of states. The S$_{11}$, S$_{22}$, and S$_{33/44}$ 
bands in Region II denote interband transitions 
in semiconducting nanotubes, while the M$_{11}$ band
corresponds to interband transitions in metallic nanotubes. The subscripts denote the sequence
of the involved Van Hove singularities with increasing energy.
The absorption bands in Region II are sitting on top of a rather
pronounced background. This background is related to an absorption
band at around 5~eV due to the collective excitations
of $\pi$ electrons in the graphene sheet.\cite{Kataura99}

\begin{figure}
\includegraphics[width=0.9\columnwidth]{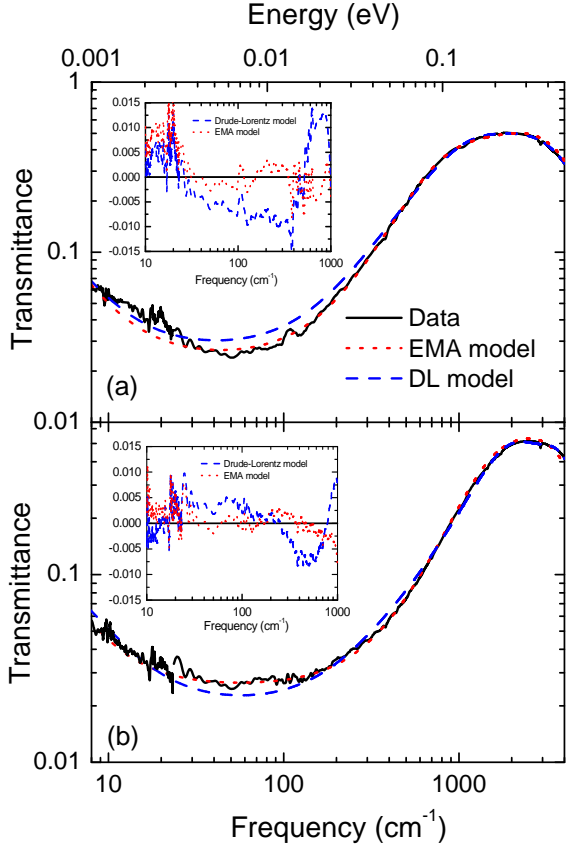}
\caption{\label{fig:RTtransfits} Room-temperature transmittance
spectra in region I (see Fig.\ \ref{fig:Ambientspectra} for
definition) of (a) the as-prepared and (b) the purified SWCNT film
together with the fits according to the Drude-Lorentz (DL) model and
the effective medium approximation (EMA) model. Insets: Difference curves
between the measured spectra and the fitting curves with the 
Drude-Lorentz model and the EMA model.}
\end{figure}

\begin{figure}
\includegraphics[width=0.9\columnwidth]{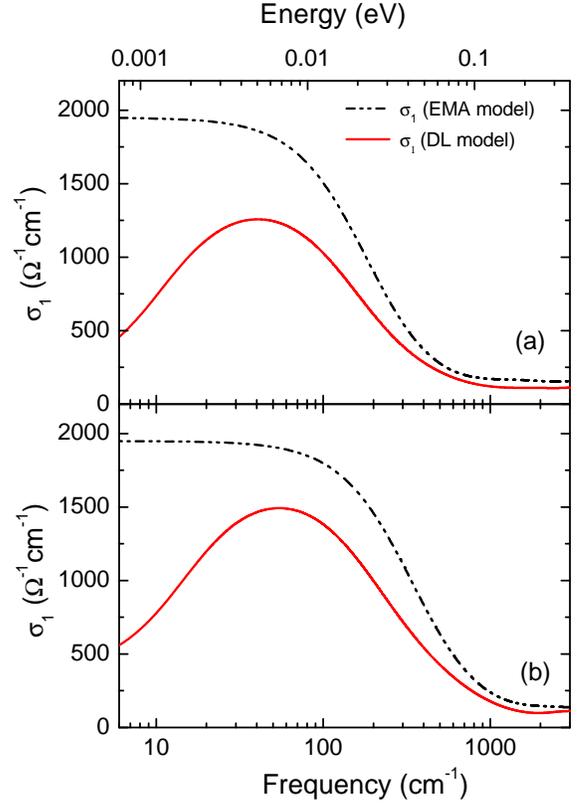}
\caption{\label{fig:sigmaRT} Low-frequency optical conductivity of
(a) the as-prepared and (b) the purified SWCNT film at ambient
conditions obtained from the fit of the transmittance spectra
using the DL model and the EMA model (see text for definitions).}
\end{figure}

\subsubsection{Low-energy excitations}
\label{sec:firambient}

The low-energy spectra of the carbon nanotubes films are dominated
by a pronounced absorption band centered in the FIR [see Region I in Fig.
\ref{fig:Ambientspectra}(b)]. For the analysis of the transmittance
spectra in this low-energy range we applied two different models:

(i) Drude-Lorentz (DL) model, which describes the excitations
of the itinerant charge carriers and the excitations of localized
carriers and interband transitions: The dielectric function of a
nanotube film according to the Drude-Lorentz model is given by
\begin{eqnarray}
\centering
\epsilon_{NTs}=\epsilon_{\infty}-\frac{\omega^{2}_{p}}{\omega^{2}+i\Gamma_{j}\omega}+\sum_{j}\frac{\omega^{2}_{p,j}}{(\omega_{j}^{2}-\omega^{2})-i\Gamma_{j}\omega}
\label{eqn:DLmodel}
\end{eqnarray}
where $\epsilon_{NTs}$ is the complex dielectric function of the
nanotube film, $\epsilon_{\infty}$ is the core contribution to the 
dielectric function, $\omega_{j}$ is the resonance frequency of the
excitation, and $\omega_{p,j}$ and $\Gamma_{j}$ are the plasma
frequency and damping of the excitation,
respectively.\cite{Dressel02} The plasma frequency $\omega_{p}$ is
defined by the expression $\omega_{p}^{2}=4\pi N e^{2}/m$ where
$N$ is the number of charge carriers and $m$ is the free-electron mass. 
The carbon nanotube film is considered as being dense enough to obtain 
the optical conductivity spectrum by applying the simple DL model 
to describe the transmittance spectrum of the carbon nanotube film 
using the coherent transmission function.\\
(ii) Maxwell-Garnett effective medium approximation (called EMA model
in the following)
combined with the DL model for finite metallic particles embedded
in a dielectric medium:\cite{Cohen73} Hereby, the nanotubes are
considered to be metallic wires of finite length embedded in a
dielectric host medium.\cite{Bommeli97, Akima06} The effective
dielectric function of the carbon nanotube film is given by the
expression
\begin{eqnarray}
\centering
\epsilon_{eff}=\epsilon_i\frac{{g+v(1-g)}\epsilon_{NTs}+(1-g)(1-v)\epsilon_i}{g(1-v)\epsilon_{NTs}+(v\cdot
g+1-g)\epsilon_i}, \label{eqn:EMAmodel}
\end{eqnarray}
where $\epsilon_i$ is the dielectric function of the insulating
dielectric host (air in our case) and $\epsilon_{NTs}$ is the
dielectric function of carbon nanotubes in the film, described by
the DL model [Eqn. (\ref{eqn:DLmodel})]. g is the geometrical
factor which determines the shape of the particles embedded in the
medium, and is equal to zero for an infinitely long needle.
Therefore, in our case where the particles are nanotubes, g tends
to zero but not exactly zero. v is the volume fraction of the
nanotubes in the dielectric medium. According to the
EMA model, the Drude
peak related to the itinerant carriers is not centered at
$\omega=0$ but at finite frequency. Within this picture, the peak
position depends more on the geometrical factor of the material in
the dielectric medium than on the carrier density.\cite{Tanner75}

The fits of the transmittance spectra of both films using the two
described models and the so-obtained optical conductivity spectra
are shown in Figs. \ref{fig:RTtransfits} and \ref{fig:sigmaRT},
respectively. Within the DL model, the FIR transmittance spectrum
can be described by two components: a Drude contribution due to
free-carriers in the metallic tubes and an absorption band described by a
Lorentzian term centered below 100~cm$^{-1}$. This gives rise to a maximum
in the FIR optical conductivity, and most of the spectral
weight lies in the band (see Fig.\ \ref{fig:sigmaRT}). Although 
a reasonable fit of the transmittance
spectra could be obtained by using this simple DL model, the
uncertainty in the fitting parameters of the Drude and the
Lorentzian terms is large.

\begin{table}[b]
\caption{Parameters obtained from the Drude fit
according to the EMA model for the as-prepared SWCNT
film (Film 1) and the purified SWCNT film (Film 2).}
\label{tab:LF-RT}
\begin{center}
\begin{tabular}{|c|c|c|}
\hline
\hline
Parameter & Film 1 & Film 2 \\
\hline
$\sigma_{dc}$ & 1949 ($\Omega^{-1}$cm$^{-1}$)& 1949 ($\Omega^{-1}$cm$^{-1}$)\\
$\omega_{p}$ & 4639 cm$^{-1}$ (139 THz)& 6345 cm$^{-1}$ (190 THz)\\
$\Gamma$ & 184 cm$^{-1}$ (5.52 THz)& 344 cm$^{-1}$ (10.31 THz)\\
\hline \hline
\end{tabular}
\end{center}
\end{table}

\begin{figure}
\includegraphics[width=1\columnwidth]{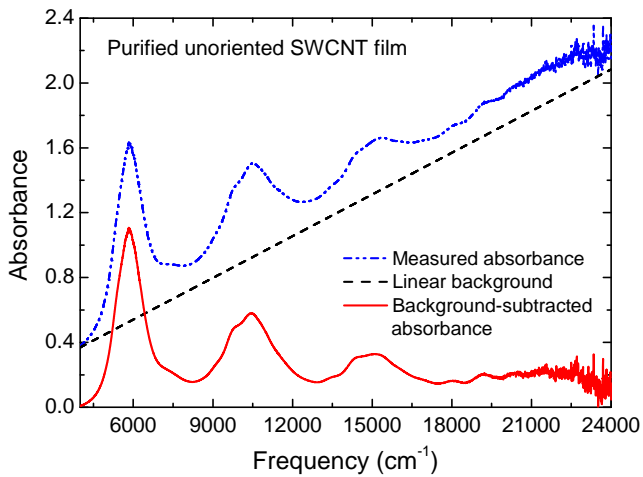}
\caption{\label{fig:bgsubtraction} Illustration of the background
subtraction procedure. The measured and the background-subtracted
absorbance spectra of the purified unoriented SWCNT film at
ambient conditions are shown together with the linear background.}
\end{figure}

Several experimental works which applied the DL model for the description
of the optical response of SWCNTs attributed the non-metallic FIR contribution, i.e.,
the band centered around 100~cm$^{-1}$, to excitations across
the curvature-induced energy gap in metallic tubes.\cite{Ugawa99,Itkis02,Borondics06,Kampfrath08} 
However, as an alternative explanation for FIR peak the resonance due 
to finite size effects arising from the morphology of
the film was proposed.\cite{Bommeli97, Jeon04, Akima06} According
to this interpretation, the EMA model should well describe the
low-frequency optical conductivity in the SWCNT films. Indeed, the
fit of the low-frequency optical response with the EMA model has 
the same quality as the fit with the DL model for both studied films 
(see insets Fig. \ref{fig:RTtransfits}). The best fit of the EMA
model was obtained for the values g=0.00025 and v=0.7. The values
of v, usually estimated based on the SEM images, lie in the
expected range of 0.6-0.9 and the value of g is much smaller than
1, as expected for the nanotube bundles.\cite{Bommeli97, Akima06}
Interestingly, it was sufficient to use only a Drude term to describe the
low-energy optical response, consistent with earlier
reports.\cite{Bommeli97, Jeon04, Akima06} Accordingly, the 
optical conductivity obtained with the EMA model comprises a purely
metallic contribution in the FIR frequency range (Fig.\
\ref{fig:sigmaRT}). Within the EMA approach the FIR
contributions arise from itinerant charge carriers that may be
localized along the tubes due to finite length effect, disorder and/or defects. 

The parameters obtained from the fit of the
EMA model are listed in Table. \ref{tab:LF-RT}. For the as-prepared film 
we found a smaller plasma frequency and a smaller scattering rate compared to
the purified SWCNT film. The smaller scattering rate might be due to the
smaller extent of the defects in the sidewalls.

\begin{figure}[t]
\includegraphics[width=0.9\columnwidth]{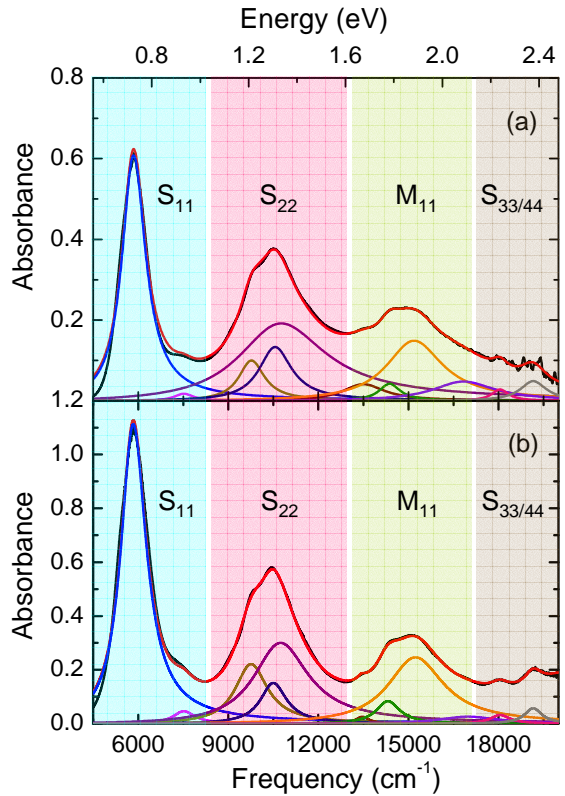}
\caption{\label{fig:FH-ambient} Background-subtracted absorbance
spectra of (a) the as-prepared and (b) the purified SWCNT film at
ambient conditions, in the near-infrared and visible frequency
ranges. The typical energy regimes of the optical transitions in
the semiconducting (marked as S) and metallic (marked as M) tubes
are represented by shaded areas of different colors.}
\end{figure}

\subsubsection{Interband transitions}
\label{sec:nir-visambient}

The energy of the interband transitions in the Region II was
extracted in the following way: The background contribution in the
MIR-to-visible part of the absorbance spectra due to $\pi$-$\pi^*$
absorption was considered as being nearly linear within the
measured energy range.\cite{Jost99, Itkis03} This linear
background was subtracted from the measured spectra, in order to
obtain only the contributions of the interband transitions. Fig.
\ref{fig:bgsubtraction} illustrates the background subtraction
procedure for one absorbance spectrum of the purified SWCNT film
as an example.
The background-subtracted spectra of both films were
then fitted using Lorentz functions to obtain the energy of the
interband transitions. Fig. \ref{fig:FH-ambient} shows the
background-subtracted absorbance spectra of the as-prepared and
the purified films at ambient conditions together with the fitting
curve and its components. The interband transition energies in both
films obtained from the fittings are tabulated in Table
\ref{tab:Eii-RT}. The results for the two 
films agree reasonably well. The fine structure observed in the
absorbance spectra is due to the contributions of nanotubes
with different diameters and chiralities.

\begin{table}[b]
\caption{Energies of the interband transitions in the as-prepared
unoriented SWCNT film (Film 1) and the purified unoriented SWCNT
film (Film 2), obtained from the fit of the background-subtracted
absorbance spectra (shown in Fig. \ref{fig:FH-ambient}). The
strength of the contributing oscillators (see text for labelling)
is marked as strong (s), medium (m) and weak (w). }
\label{tab:Eii-RT}
\begin{center}
\begin{tabular}{|c|c|c|}
\hline \hline
\multirow{2}{*}{Label} & \multicolumn{2}{|c|}{Frequency (cm$^{-1}$)} \\
 & Film 1 & Film 2 \\
\hline
S$_{11}$ (s) & 5843 & 5834 \\
S$_{11}$ (w) & 7516 & 7516 \\
S$_{22}$ (m) & 9771 & 9759 \\
S$_{22}$ (m) & 10560 & 10500 \\
S$_{22}$ (s) & 10764 & 10742 \\
M$_{11}$ (w) & 13500 & 13438 \\
M$_{11}$ (w) & 14372 & 14319 \\
M$_{11}$ (s) & 15190 & 15235 \\
M$_{11}$ (w) & 16834 & 16982 \\
S$_{33/44}$ (w) & 18053 & 18053 \\
S$_{33/44}$ (w) & 19157 & 19157 \\
\hline \hline
\end{tabular}
\end{center}
\end{table}

\begin{figure}[t]
\includegraphics[width=0.9\columnwidth]{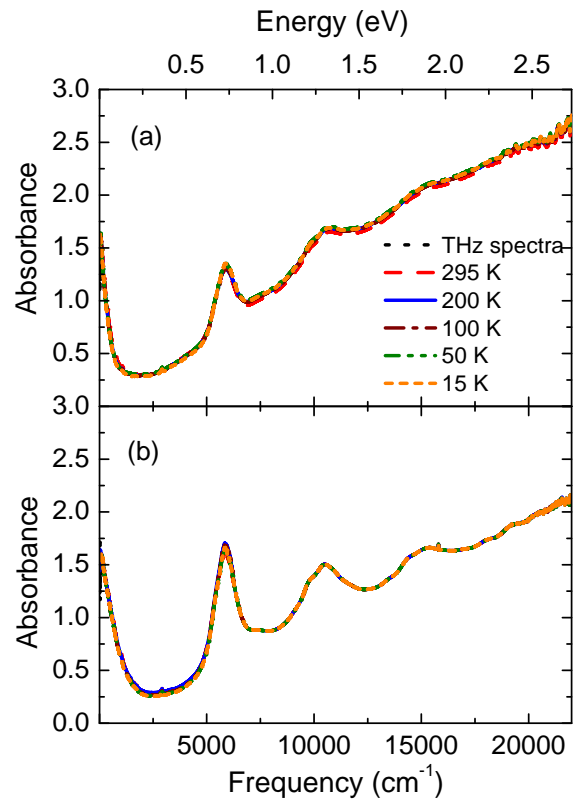}
\caption{\label{fig:FH-Tplot} Absorbance spectra of (a) the
as-prepared and (b) the purified SWCNT film as a function of
temperature over a broad frequency range (6-22000~cm$^{-1}$).}
\end{figure}

\subsection{\label{sec:T-dependence}Temperature dependence}

The temperature-dependent absorbance spectra of the
as-prepared and the purified SWCNT film are depicted in
Fig. \ref{fig:FH-Tplot}. The spectra for both films are 
very similar and nearly temperature-independent
over the entire measured frequency range (6-22000~cm$^{-1}$). 
In earlier temperature-dependent infrared spectroscopic
measurements on unoriented SWCNTs an extremely weak non-metallic
temperature dependence was observed in the temperature range from
8~K to 298~K.\cite{Ugawa99,Ugawa01} A more recent
temperature-dependent study on carbon nanotube networks by
Borondics \emph{et al.} also revealed only weak changes with respect
to temperature in the FIR spectral regime.\cite{Borondics06} 
This weak temperature dependence in the FIR absorption can be 
attributed to a tube-to-tube variation of the chemical potential.\cite{Kampfrath08}
Thus, our finding of the negligible temperature dependence of the
studied as-prepared and purified SWCNT films, is in good
agreement with earlier investigations.

\subsection{\label{sec:P-dependence}Pressure dependence}

\begin{figure}
\includegraphics[width=0.9\columnwidth]{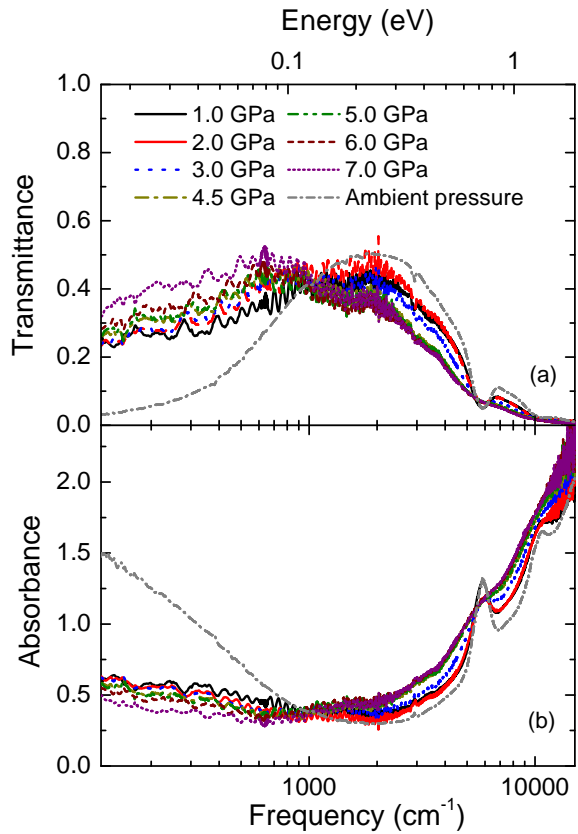}
\caption{\label{fig:FH6-Pplot} (a) Transmittance and (b)
absorbance spectra of the as-prepared SWCNT film for selected
pressures up to 7~GPa over a broad frequency range
(120-15000~cm$^{-1}$).}
\end{figure}

\begin{figure}
\includegraphics[width=0.9\columnwidth]{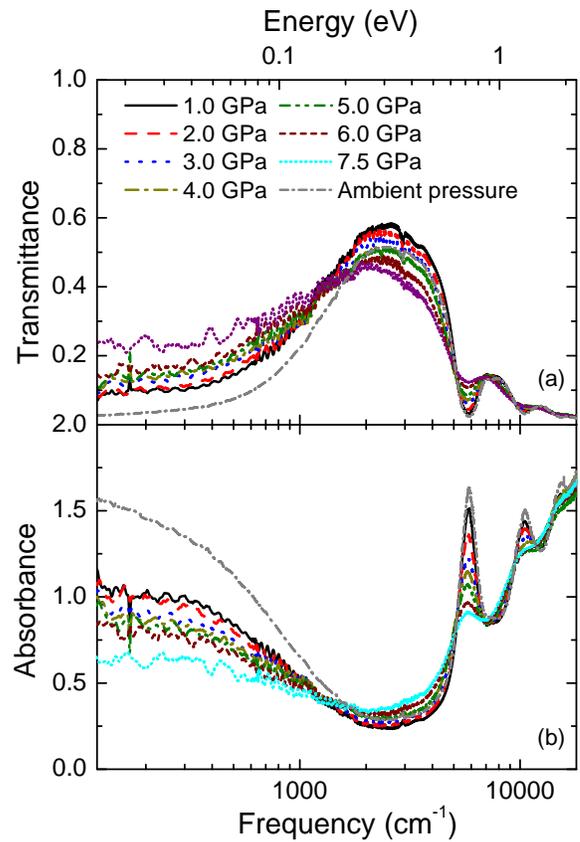}
\caption{\label{fig:FH5-Pplot} (a) Transmittance and (b)
absorbance spectra of the purified SWCNT film for selected
pressures up to 7.5~GPa over a broad frequency range
(120-20000~cm$^{-1}$).}
\end{figure}

The as-prepared and the purified SWCNT films exhibit considerable
changes in the electronic properties with the application of
pressure in both the Regions I and II. Figs. \ref{fig:FH6-Pplot}
and \ref{fig:FH5-Pplot} show the transmittance and absorbance
spectra of the as-prepared and the purified unoriented SWCNT films
for selected pressures up to 8~GPa in the FIR-to-visible frequency
range. With increasing pressure, the interband transitions of the
carbon nanotubes shift to lower frequencies and also exhibit a
significant broadening. The intensity of the low-energy absorption
($<$1000~cm$^{-1}$) decreases with increasing pressure. In the
following we will first present the pressure dependence of the
low-energy excitations, and then focus on the interband transitions
as a function of pressure.

\subsubsection{Low-energy excitations}

\begin{figure}
\includegraphics[width=0.9\columnwidth]{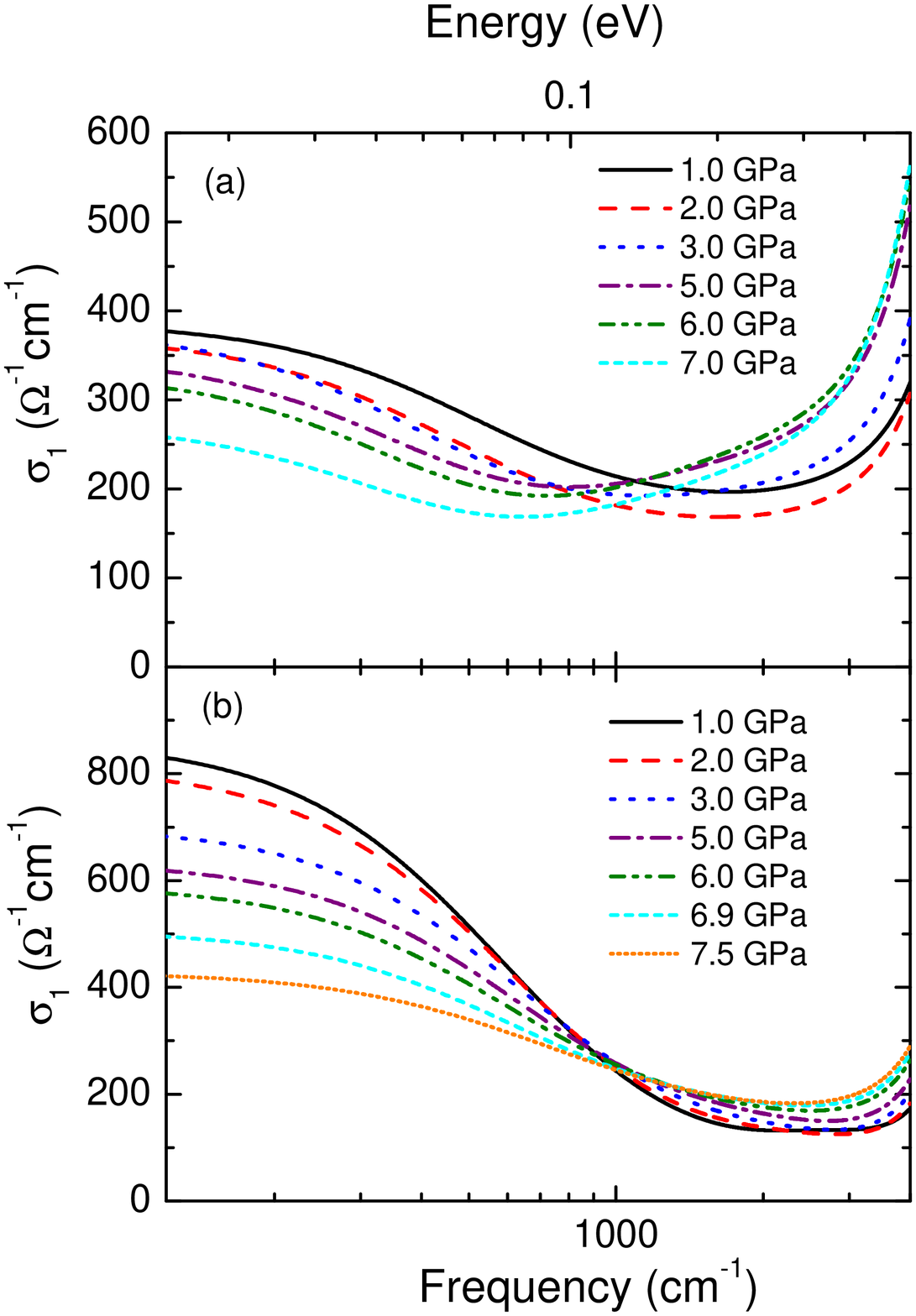}
\caption{\label{fig:sigvsP} Pressure-dependent optical
conductivity spectra of (a) the as-prepared and (b) the purified
SWCNT film obtained from the fit of transmittance spectra using
the EMA model (see text for definition).}
\end{figure}

The low-energy absorbance spectra of the carbon nanotubes were
found to be significantly affected by the application of pressure
(see Figs. \ref{fig:FH6-Pplot} and \ref{fig:FH5-Pplot}). In Figs.
\ref{fig:FH6-Pplot} and \ref{fig:FH5-Pplot} we also include the
ambient-pressure spectra measured on the free-standing samples.
The absorbance of the SWCNT films at the lowest measured pressure
(1~GPa) is significantly lower compared to that at ambient
pressure, with the effect being more drastic for the as-prepared
SWCNT film. It seems that the adsorption of the pressure medium
strongly influences the itinerant carriers.

\begin{figure}
\includegraphics[width=0.9\columnwidth]{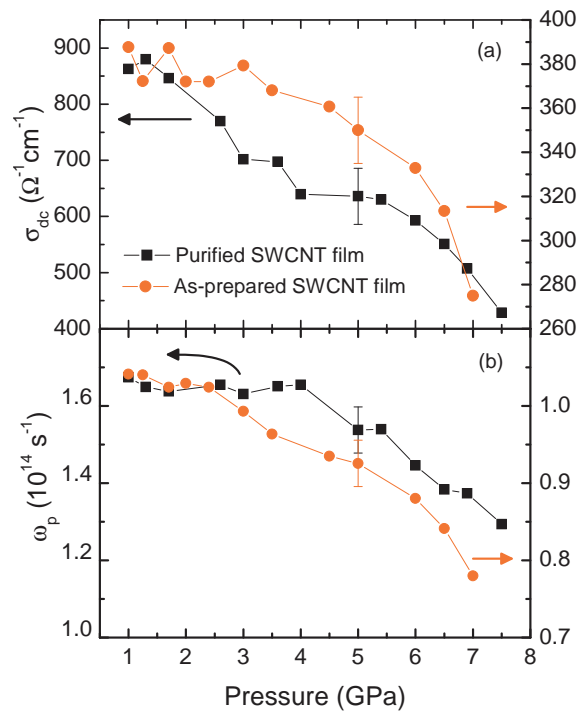}
\caption{\label{fig:drudeterms} Pressure dependence of (a) the dc
conductivity $\sigma_{dc}$ and (b) the plasma frequency
$\omega_{p}$ of the Drude term in the FIR optical conductivity of
both studied SWCNT films obtained from the fit using the EMA model
(see text for definition).}
\end{figure}

With increasing pressure the absorbance of the SWCNT films decreases
continuously. For a quantitative
analysis of the pressure-induced changes in the low-energy
infrared response of the carbon nanotubes, the transmittance
spectra were fitted with the EMA model as described in Section
\ref{sec:firambient}. The parameters of the EMA model, namely, the
geometrical factor and the volume fraction, were assumed 
to be pressure-independent, in order to minimize the
uncertainties in the analysis. This is also a reasonable assumption,
since these parameters should not be changed with the application of
external pressure. Similar to the ambient-pressure spectra, the
low-frequency part of the spectra can be described by one Drude
contribution. Fig. \ref{fig:sigvsP} shows the so-obtained optical
conductivity spectra of the as-prepared and the purified SWCNT
films for various pressures up to 7.5~GPa. The low-frequency
optical conductivity decreases with increasing pressure, in good
qualitative agreement with the absorbance spectra. The dc
conductivity $\sigma_{dc}$ and the plasma frequency
$\omega_{p}$ of the Drude term obtained from the fitting
decrease with increasing pressure (see Fig. \ref{fig:drudeterms}).
The pressure dependence of both films are in qualitative
agreement.

\subsubsection{Interband transitions}

\begin{figure}
\includegraphics[width=1\columnwidth]{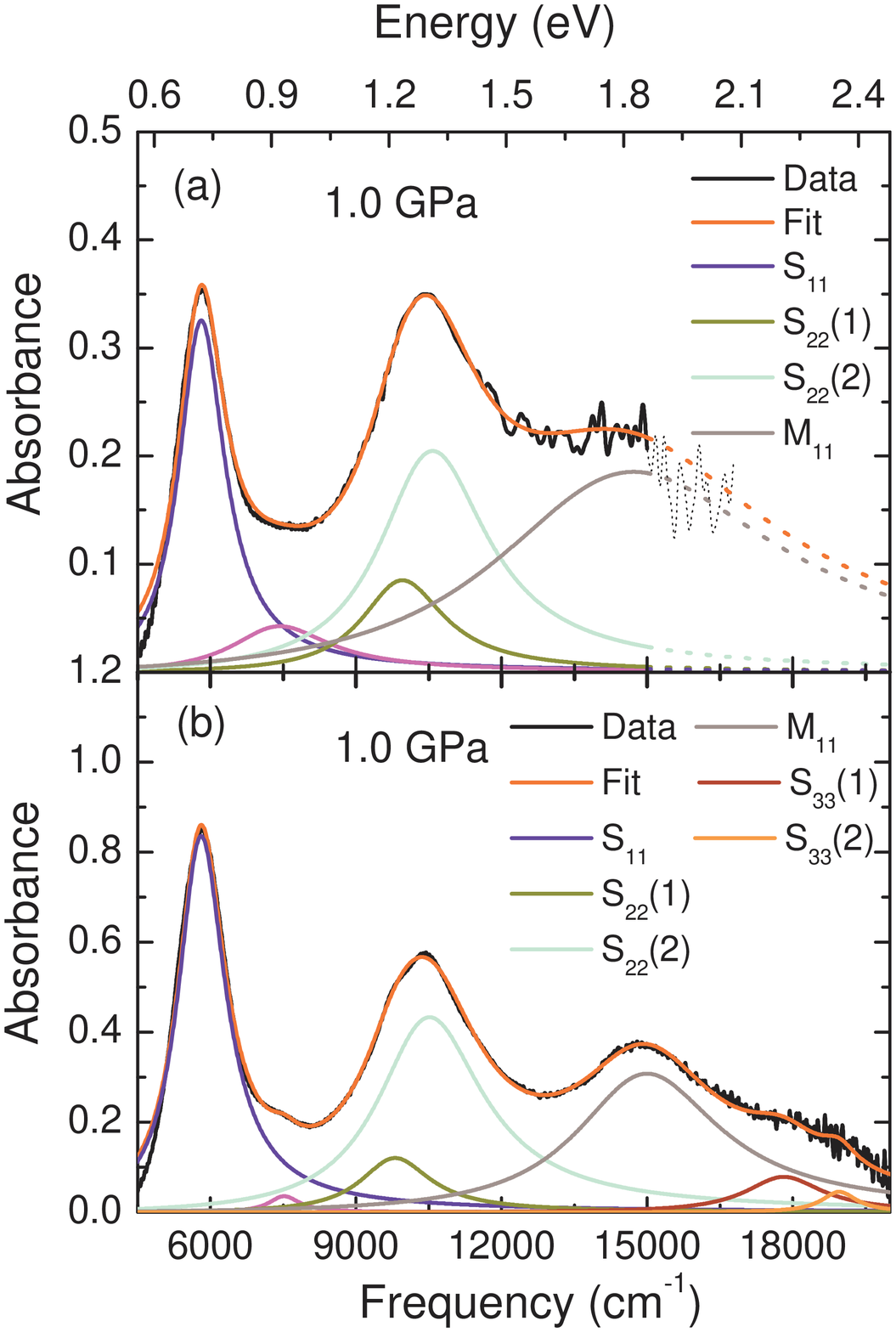}
\caption{\label{fig:FH-LPfitterms} Background-subtracted
absorbance spectra of (a) the as-prepared and (b) the purified
SWCNT film at lowest pressure together with the fit of the optical
transitions using Lorentzian oscillators. Only the strong
transitions are considered and labeled.}
\end{figure}

For a quantitative analysis of the pressure-induced changes in the
interband transitions their energies were extracted from the
absorbance spectra for all measured pressures using the same
procedure as for the absorbance spectra at ambient conditions
(i.e., background subtraction and fitting with Lorentzian functions, see
Section \ref{sec:nir-visambient}). For an illustration of the procedure 
we show in Fig.\ \ref{fig:FH-LPfitterms} the background-subtracted
absorbance spectra of the as-prepared and the purified SWCNT films
for the lowest measured pressure, together with the fitting curve
and its components. It is obvious that the fine structure in the
optical absorption bands observed in the spectrum at ambient
pressure (see Fig.\ \ref{fig:FH-ambient}) is obscured even at low
pressures by pressure-induced broadening effects. For the same
reason, only the strong and most obvious transitions have been
considered for the fitting using Lorentzian functions, and
excellent fits were accomplished.

\begin{figure}
\includegraphics[width=1\columnwidth]{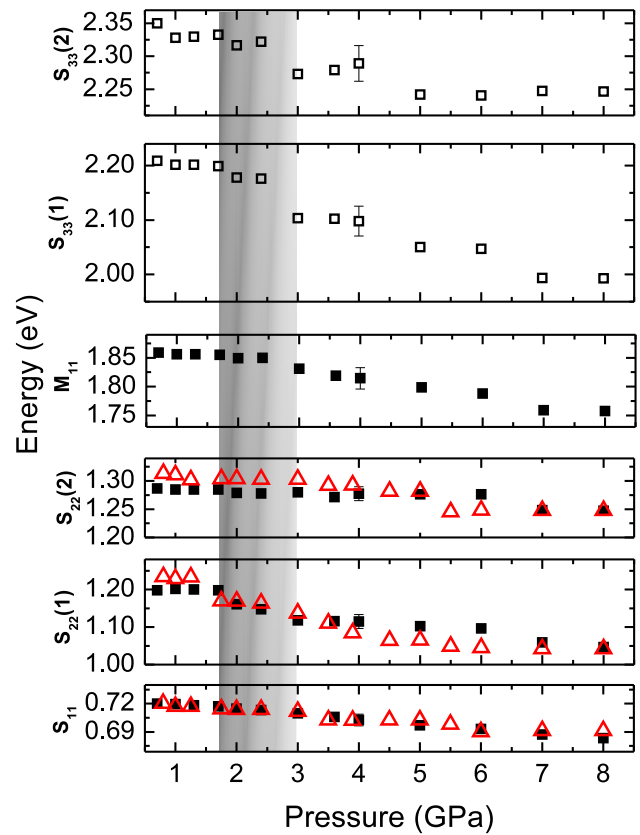}
\caption{\label{fig:eii-fh} Energies of the strong optical
transitions (labeled transitions in Fig. \ref{fig:FH-LPfitterms})
of the as-prepared SWCNT film (hollow triangles) and the purified
SWCNT film (solid squares), corresponding to semiconducting
(marked as S) and metallic (marked as M) nanotubes as a function
of pressure up to 8~GPa.}
\end{figure}

The so-obtained energies of the interband transitions with the 
strongest contribution (labeled in Fig.\ \ref{fig:FH-LPfitterms}) 
of both SWCNT films are shown in Fig.~\ref{fig:eii-fh} as a function
of pressure. In case of the as-prepared SWCNT film only the first two
interband transitions could be unambiguously traced as a function of
pressure due to the low signal-to-noise ratio above 15000~cm$^{-1}$.
The pressure dependence of the interband transitions in both SWCNT
films are in good agreement and the important observations are as
follows: (i) All interband transitions exhibit a small redshift with 
increasing pressure up to the critical pressure $P_c$$\approx$2.0~GPa.
(ii) Above $P_c$ the redshift of the interband transitions is enhanced.
This results in an (iii) anomaly in the pressure-induced shift at $P_c$.

\section{Discussion} \label{sec:discussion}

\subsection{Low-energy excitations}

The FIR conductivity of the SWCNT films is nearly independent of temperature.
The complex dielectric constant measurements by Hilt
\textit{et al}.,\cite{Hilt00} indicate that the transport
properties of the SWCNT mats can be well described by a Drude
model with a negligible temperature dependence. They observed that
the conductivity at very low frequencies (around 285~GHz) is
dominated by localized carriers. Based on temperature-dependent
measurements on unoriented carbon nanotubes, including
nanotubes embedded in the polymer matrix, is was concluded that
there is a localization of carriers in the bundles and along
the tubes.\cite{Hilt00, Benoit02, Fuhrer99} Additionally, theoretical
studies found that the localization of carriers plays a
major role in determining the low-energy properties of carbon
nanotubes.\cite{Jiang01,Hjort01} Hence, carrier localization effects
should be considered in the discussion of our results. 

The application of pressure strongly influences the Drude conductivity 
of the as-prepared and the purified SWCNT films:
The dc optical conductivity and the plasma frequency decrease with 
increasing pressure (see Fig.\ \ref{fig:drudeterms}). 
Hence, pressure enhances the localization effects of the charge
carriers. Recent theoretical investigations on the
localization effects on deformed nanotubes indeed suggest that with
increasing deformations, the localization length of the carriers
decreases leading to a decrease in the conductivity.\cite{Bu02,Hjort01}
In addition to the localization effects, the electronic band
structure of the nanotubes is expected to change considerably
under pressure due to structural changes in the nanotubes. The
deformation of the nanotubes induces symmetry breaking and
$\sigma^*$-$\pi^*$ hybridization effects due to the mixing of
states in the polygonized nanotubes. A possible interaction between the
facing layers in the radially-deformed nanotubes further modifies
the low-energy electronic structure strongly.\cite{Park99,Yang00,Lammert00} 
These changes eventually lead to the
opening or closing of the gap at the Fermi level and thereby
causing a metal-to-semiconductor transition or vice-versa
in metallic tubes or small-gap nanotubes,
respectively.\cite{Charlier96, Park99, Mazzoni00, Lammert00,
Gulseren02, Lu03} Although this could lead to competing effects
on the low-energy conductivity of SWCNT films, we
observe an overall decrease in the optical conductivity with increasing pressure. 
It seems that only a small fraction of the small-gap nanotubes in our sample undergo
a semiconductor-to-metal transition under pressure.

Therefore, we tentatively assign the pressure-induced decrease in the low-energy 
conductivity to the increasing localization of carriers caused by defects and 
deformations combined with the decrease in the density of 
states close to the Fermi level due to the radial deformation of the SWCNTs. 
This interesting observation calls for further experimental investigations
of the pressure-dependent electronic properties of carbon nanotubes down 
to lower frequencies. It is important to note that the Drude parameters
(dc conductivity, plasma frequency, see Fig.\ \ref{fig:drudeterms}) obtained 
from the analysis of our data have relatively large error bars, since it is not 
possible to study the frequency range below $\approx$100~cm$^{-1}$ with our 
technique due to the diffraction limit. This could explain why we do not find 
an anomaly in the pressure dependence of these parameters at $P_c$.

\subsection{Interband transitions}

Pressure and temperature affect the interband transitions of
the carbon nanotubes in a very different way. Although the
electronic properties of both the as-prepared and the purified
films show considerable changes with the application of pressure,
the two SWCNT films exhibit temperature-independent properties.
The temperature-dependent energy shifts of the interband transitions
of the carbon nanotubes in Region II, are expected to be either
positive or negative depending on the chiral angle in the
photoluminescence measurement.\cite{Li04} Recent theoretical
studies found that the temperature-dependence of the bandgap of
the carbon nanotubes is relatively small compared to the bulk
semiconductors, and the maximum temperature-induced shift was
calculated to be around 10~meV.\cite{Capaz05} In addition to the
small temperature-induced changes which are highly chirality-dependent, 
it was suggested that the temperature-dependence of
optical transition energies of the nanotubes in the bundles are
dominated by the thermal expansion of the nanotubes'
environment.\cite{Cronin06} Very recent experimental and
theoretical temperature-dependent studies on semiconducting
nanotubes suggest that the positive or negative energy shifts
could be attributed to external strain, while the broadening of
the transitions might be attributed to electron-phonon
interactions.\cite{Karaiskaj07} These effects are not relevant
here, since the nanotube films studied within
this work are free-standing films. The interband transitions
of the unoriented and unsegregated nanotube films are generally
convolutions of the transitions corresponding to tubes with
different diameters and chiralities. These transitions are
therefore quite broad when compared to the small changes expected
during the change in temperature. Thus, the observation of an
``averaged-out'' temperature change in the nanotube films which
give rise to the temperature-independent NIR-visible frequency
absorbance spectra is plausible.

The changes in the optical transition energies under hydrostatic
pressure were theoretically investigated in Refs.\ \cite{Capaz04,Liu08} 
A ``family behavior'' in the pressure
coefficients, i.e., having positive or negative values of pressure
coefficients depending on chirality, was found for
semiconducting SWCNTs. A similar behavior was reported for the 
changes of the band gap under uniaxial stress.\cite{Yang99,
Yang00, Gartstein03} On the contrary, experimentally an overall shift of
the interband transitions to lower energies with increasing pressure
was observed in both bundled and individualized
nanotubes.\cite{Kazaoui00, Wu04, Deacon06} The shift of the
interband transitions to lower energies was attributed to
the intertube interactions and/or symmetry breaking in the very
early high-pressure optical absorption
measurements.\cite{Kazaoui00} Photoluminescence measurements
on individualized nanotubes in aqueous solution of surfactant
demonstrated that the negative pressure coefficient is an intrinsic
property of the individual nanotubes and the intertube
interactions do not play a role.\cite{Wu04} Furthermore, 
$\sigma^*$-$\pi^*$ hybridization was suggested as an explanation 
for the overall negative pressure coefficient, at least in the 
low-pressure regime (below 1.3~GPa).\cite{Wu04}

In an alternative approach, Deacon \textit{et al.} attributed the
downshift of the interband transitions to the interaction of the 
nanotubes with the surrounding surfactant:\cite{Deacon06}  
The dielectric constant of the surfactant solution (which is the
surrounding medium) increases with increasing pressure, causing
a reduction of the Coulomb interactions. Also a recent
investigation on individual double-walled nanotubes under pressure
showed that the interaction of nanotubes with the surrounding
medium is very important.\cite{Puech06} The pressure coefficients
of the Raman bands were highly dependent on the employed pressure
transmitting medium.\cite{Puech06} However, the
lowering of the energy of the interband transitions was
also observed in the case of a solid pressure transmitting medium,
for which adsorption of the pressure transmitting medium can be
ruled out.\cite{Kazaoui00, Kuntscher08} Therefore, the decrease
in energy of the interband transitions with increasing pressure cannot 
be solely attributed to the effect of the surrounding medium,
and thus hybridization effects have to be considered.

The $\sigma^*$-$\pi^*$ hybridization not only plays a significant
role in the electronic band-structure of the small diameter tubes
\cite{Blase94} but also in polygonized and deformed tubes. The
changes in the electronic properties of the SWCNTs with
polygonized cross-sections were theoretically
investigated.\cite{Charlier96, Liu08} According to these studies,
the radial deformation of the nanotubes strongly influences the
band-structure and hence the optical properties of the nanotubes. 
It was found that the deformation-induced hybridization effects 
and symmetry-breaking lower the conduction-band states towards the
Fermi level,\cite{Charlier96,Liu08} which nicely explains the 
redshift of the interband transitions observed in our pressure-dependent
optical data.

\begin{figure}
\includegraphics[width=1\columnwidth]{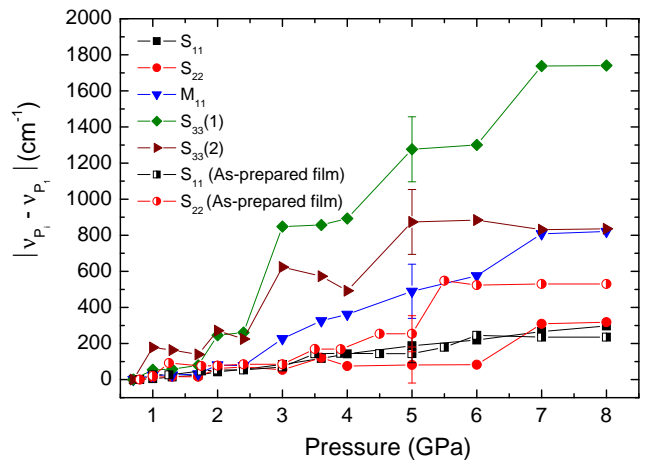}
\caption{\label{fig:deltanu} Pressure-induced shift of the interband transitions 
with respect to the lowest pressure, calculated as the difference between
the energy of the transition, $\nu_{P_i}$, at a certain pressure $P_i$, and the 
the corresponding energy at the lowest applied pressure, $\nu_{P_1}$.
The labels of the interband transitions correspond to that of the Fig.\ \ref{fig:eii-fh}.
The transition energies for the as-prepared SWCNT film are indicated.}
\end{figure}

The pressure-induced shift of the interband transitions with respect to 
the lowest applied pressure is depicted in Fig.\  \ref{fig:deltanu}. 
The results for both films are very similar. Compared to the lowest-energy
transition S$_{11}$, the transitions in semiconducting tubes at higher
energies, namely S$_{22}$ and S$_{33}$, and those in metallic
tubes, M$_{11}$, are more sensitive regarding pressure above 2~GPa
compared to the first transition S$_{11}$. According to earlier
pressure-dependent optical absorption studies \cite{Kazaoui00}
the interband transitions of semiconducting tubes
vanish at a pressure of 4.1~GPa, while the transitions
corresponding to the metallic tubes persist. In contrast, we observe that
the pressure dependence of the interband transitions in
semiconducting and metallic tubes are very similar, with comparable
pressure coefficients (see Fig.\ \ref{fig:deltanu}). The overall
red-shift of the interband transitions is in qualitative agreement
with earlier pressure-dependent studies on
SWCNTs.\cite{Kazaoui00,Wu04} However, we observe much smaller energy shifts 
than those in earlier reports.\cite{Kazaoui00} 
For example, the shift of the energy of S$_{11}$ transition in the 
measurements by Kazaoui \emph{et al.} at 4~GPa is nearly twice as much 
as that found in our data. This discrepancy is probably
due to the less hydrostatic conditions present during previous 
experiments. In our experiment we used a more hydrostatic pressure
medium (argon), which minimizes strains due to non-hydrostaticity,
thus reducing the extent of the red-shift and the broadening of the
observed absorption bands.

On applying pressure, the interband transitions show small
shifts up to $P_c$$\approx$2~GPa. Above this critical pressure, the energy of the
interband transitions decrease more drastically with increasing
pressure, causing an anomaly in the pressure dependence at $P_c$.
Several Raman spectroscopic investigations also reported the
disappearance or loss of intensity of the radial breathing mode
(RBM) of nanotubes and anomalies in the tangential mode of
nanotubes at pressures in the range 1.5-2~GPa, which were attributed to the
radial deformation of nanotubes.\cite{Loa03}

There is an enormous number of theoretical studies addressing the
pressure-induced structural deformations in carbon nanotubes. The
circular cross-section of the nanotubes undergoes significant
deformation under pressure.\cite{Okada01, Reich02, Chan03} It was
suggested that the circular cross-section of nanotubes
deforms and stabilizes at an oval, elliptical, racetrack-like or
peanut-shaped cross-section.\cite{Reich02, Chan03, Sluiter04,
LiChou04, Zhang04, Elliott04, Capaz04} The nature of the phase
transition and the deformation in the carbon nanotubes are very
complex with the existence of metastable states, and are highly
dependent on the nanotube diameter just like the other properties
of nanotubes.\cite{Sluiter04, Elliott04, Hasegawa06} 
Despite some discrepancies concerning the details of the high-pressure
phase of the carbon nanotubes, it is well established through both
experimental and theoretical studies that the discontinuous radial
deformation in SWCNTs occurs at a certain critical pressure P$_c$.
The value of P$_c$ decreases with increasing carbon nanotube
diameter and follows the relation P$_c$ $\propto$ 1/d$^3$ for both
individual SWCNTs\cite{Capaz04, Zhang04, LiChou04, Hasegawa06} and
bundled SWCNTs.\cite{Zhang04, Elliott04, LiChou04, Capaz04} For
nanotubes grown by the laser ablation technique, with an average
diameter of 1.2-1.4~nm, the critical transition pressure is
expected to be about 2~GPa.\cite{Chan03, Sluiter04, Elliott04,
Capaz04, Hasegawa06} Hence, our experimental finding of an anomaly
at P$_c$ $\approx$ 2~GPa is in agreement with the above-mentioned
earlier reports. Furthermore, a recent theoretical investigation on
the nature of the deformation of the SWCNTs found that the SWCNTs
of diameter less than 2.5~nm undergo a transition from a circular
to an oval shape and do not collapse, while the tubes with larger
diameters do collapse.\cite{Hasegawa06} Hence, the anomaly found
in our experiments could be attributed to the structural phase
transition in the nanotubes, where the circular cross-section is
deformed to an oval shape.

Resonant Raman studies on individualized nanotubes under pressure
report the critical pressures 10~GPa and 4~GPa
for radial deformation for nanotubes of diameters 0.8-0.9~nm and 1.2-1.3~nm,
respectively.\cite{Lebedkin06} Furthermore, it is interesting to
note that some experiments found no significant deformation
effects up to high pressures in bundled nanotubes.\cite{Amer04,
Merlen05, Merlen06, Monteverde05}  For example, no anomaly was
observed by resonant Raman spectroscopy on carbon nanotubes
for pressures up to 40~GPa.\cite{Merlen05} The absence of an
anomaly was attributed to the adsorption of argon in the
tubes.\cite{Merlen05} Other recent measurements on purified
nanotubes also suggest that the pressure transmitting medium plays a
decisive role for the critical pressure of the
the structural phase transition, due to adsorption
effects.\cite{Amer04, Proctor06} In contrast to these reports, we
do observe an anomaly in the pressure dependence of the optical
transitions when using argon as pressure transmitting medium.
Our observation of an anomaly does, however, not exclude the
physical adsorption of argon on the SWCNTs during our pressure
measurements.

Although the electronic properties of bundled carbon nanotubes
are changed considerably with increasing pressure, the
pressure-induced changes are almost completely reversible: The
position of the interband transitions were reversible on releasing
pressure, but the intensities of the interband transitions were
not regained completely. The reversibility of the pressure-induced
transformations in the nanotubes has been
controversial.\cite{Loa03, Sluiter04} Sluiter et al., suggested a
phase diagram for SWCNT bundles which could explain the
inconsistency in the reversibility of the pressure-induced changes
in various experimental investigations.\cite{Sluiter04} According
to this phase diagram, the critical pressure at which the radial
deformation occurs is lower than the cross-linking pressure. At
the cross-linking pressure, the deformed nanotubes in the bundles
start linking to each other similar to a polymerization effect.
Therefore, the reversibility of the experimental data may not be
observed when the highest measured pressure is above the
cross-linking pressure. Despite the observed reversibility of the
pressure-induced effects in our data, the crosslinking of the nanotubes
cannot be completely ruled out in our experiments: 
Density functional electronic structure simulations predict that 
the cross-linking is reversible when the circumference of the nanotubes 
is greater than 3.8~nm.\cite{Sluiter04} The SWCNTs in our work have an 
average circumference of 4.1~nm, which implies that the cross-linking,
if it occurs, would be completely reversible upon pressure release.

\section{Conclusions} \label{sec:conclusions}

We have investigated the optical properties of as-prepared and purified 
unoriented SWCNT films over a broad frequency range (FIR-to-visible) as a 
function of temperature and pressure (up to 8~GPa). Both the as-prepared and the
purified SWCNT film exhibit nearly temperature-independent
properties. With increasing pressure the low-energy absorbance monotonically 
decreases due to enhanced carrier localization. The energy of the
interband transitions decreases with application of pressure due to the significant 
hybridization and symmetry-breaking effects. At around 2~GPa the shifts of the 
interband transitions show an anomaly, which can be attributed to a pressure-induced
structural phase transition.

\begin{acknowledgments}
We thank A. Abouelsayed and M. Kappes for fruitful discussions.
We acknowledge the ANKA Angstr\"omquelle Karlsruhe for the provision of beamtime
and thank B. Gasharova, Y.-L. Mathis, D. Moss, and M. S\"upfle for assistance at the 
beamline ANKA-IR. We thank M. Dressel for the opportunity to perform THz frequency
range measurements at the University Stuttgart, 1. Physikalisches Institut, 
and F. Rauscher and K. Venkataramani for their help during the THz measurements.
We also thank D. Trojak for her support regarding the SEM images of the studied 
nanotube films. Financial support by the Deutsche Forschungsgemeinschaft (KU 1432/3-1)
and the Hungarian Academy of Sciences (grant No.\ DFG/183) is gratefully 
acknowledged.
\end{acknowledgments}

\end{document}